\documentclass{article}[10pt]
\usepackage{graphicx}
\def\Journal#1#2#3#4{{\it #1} {\bf #2}, #3 (#4)}

\makeindex
\begin{document}

\title{\bf MICE: The International Muon Ionization Cooling Experiment\footnotemark}

\author{D. M. Kaplan\footnotemark~$^{,1}$ and K. Long$^{2}$\\
\it $^{1}$Physics Division, Illinois Institute of Technology, 
\it Chicago, Illinois 60616, USA\\[0.05in]
\it $^{2}$Imperial College London, UK\\[0.1in]
for the MICE Collaboration\footnotemark}

\twocolumn
[\begin{small} \maketitle\abstract{
Muon storage rings have been proposed for use as a source of high-energy neutrino beams (the Neutrino Factory) and as the basis for a high-energy lepton-antilepton collider (the Muon Collider).  The Neutrino Factory is widely believed to be the machine of choice for the search for leptonic {\em CP} violation while the Muon Collider may prove to be the most practical route to multi-TeV lepton-antilepton collisions.  The baseline conceptual designs for each of these facilities requires the phase-space compression (cooling) of the muon beams prior to acceleration.  The short muon lifetime makes it impossible to employ traditional techniques to cool the beam while maintaining the muon-beam intensity.  Ionization cooling, a process in which the muon beam is passed through a series of liquid-hydrogen absorbers followed by accelerating RF cavities, is the technique proposed to cool the muon beam.  The international Muon Ionization Cooling Experiment (MICE) collaboration will carry out a systematic study of ionization cooling.  The MICE experiment, which is under construction at the Rutherford Appleton Laboratory, will begin to take data late this year.  The MICE cooling channel, the instrumentation and the implementation at the Rutherford Appleton Laboratory are described together with the predicted performance of the channel and the measurements that will be made.
}
\vspace{0.4in}\end{small}]

\renewcommand{\thefootnote}{\fnsymbol{footnote}}
\footnotetext[1]{Contributed to XXIII International Symposium
on Lepton and Photon Interactions at High Energy (LP07), Daegu, Korea, Aug.\ 13--18, 2007.}
\footnotetext[2]{E-mail address: kaplan@iit.edu}
\footnotetext[3]{Website: {\tt http://www.mice.iit.edu/}}
\renewcommand{\thefootnote}{\arabic{footnote}}

\section{Introduction}
Ionization cooling~\cite{cooling}, a key enabling technology for intense stored muon beams, was proposed over 20 years ago, but its practicality has yet to be shown. Applications include: (1)~neutrino factories, possibly the ultimate tool for the study of neutrino oscillation and leptonic CP violation~\cite{Lindner}; and (2)~muon colliders, whose low levels of beamstrahlung and synchrotron radiation can enable precision studies of $s$-channel-produced Higgs bosons, as well as  high-luminosity multi-TeV lepton-antilepton collisions~\cite{Kaplan-ICHEP}. 

MICE is an experimental program to verify the feasibility and performance of ionization cooling: the tunable emittance of muon beams from 140 to 240\,MeV/$c$ will be precisely measured before and after a section of ionization-cooling channel. The MICE collaboration comprises some 140 accelerator and particle physicists and engineers from Belgium, Bulgaria, China, Italy, Japan, the Netherlands, Switzerland, the UK and the US. The experiment is approved and under construction at the Rutherford Appleton Laboratory (RAL) in the UK.

\begin{figure}[h]
\centerline
{\includegraphics[width=\linewidth,bb=22 85 749 460,clip]{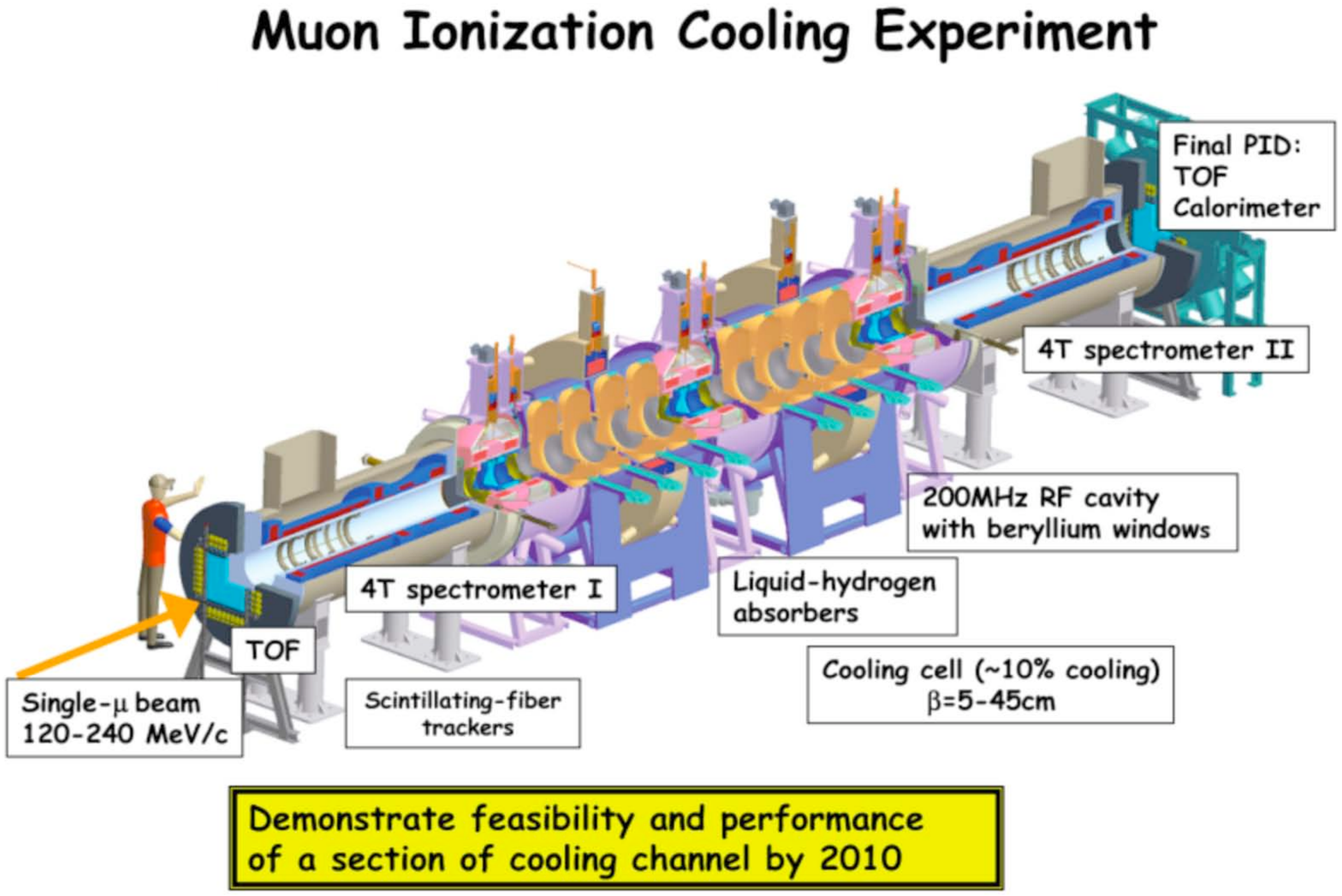}}
\vspace{-.08in}
\caption{3D rendering of MICE.}
\vspace{-.13in}
\label{fig:MICE}
\end{figure}

\section{Experiment Description}

The muon beam is produced by 800 MeV protons from RAL's ISIS synchrotron and momentum-selected and transported by the MICE beamline to the apparatus shown in  Fig.~\ref{fig:MICE}. Particle identification (PID) ensures muon purity better than 99.9\%. The input beam is tunable from 1 to 12$\pi$\,mm$\cdot$rad input emittance. The 6D emittance is measured in a 5-station scintillating-fiber tracker immersed in the 4\,T uniform magnetic field of a superconducting solenoid. The tracker determines $x,x^\prime,y,y^\prime$, and particle energy, while time-of-flight (TOF) counters provide the sixth phase-space coordinate, $t$. 

The cooling section is a series of absorbers and normal-conducting RF cavities, with superconducting coils providing an 
axisymmetric  focusing field. The exiting beam emittance is measured in a second spectrometer system (tracker and TOF) identical to the first one. Electrons from muon decay would bias the emittance measurement and are removed via a calorimeter.

\section{Technical Challenges of\\ MICE}

Despite its conceptual simplicity, ionization cooling poses several challenges:
\begin{enumerate}
\item Operating high-gradient (16\,MV/m) RF cavities of relatively low frequency (201\,MHz) in strong  magnetic fields (1--3\,T). As shown in the MuCool program at Fermilab, this can produce intense dark currents, heating the LH$_2$ absorbers or causing cavity breakdown.
\item Designs for safe operation with substantial amounts of LH$_2$ near RF cavities.
\item The small cooling effect ($\approx$10\%) from an affordable device sets the goal of $10^{-3}$ emittance precision, achieved via a highly segmented, low-mass, precise tracker with low multiple scattering and high redundancy against dark-current-induced background. 
\end{enumerate}

\section{Status of the Project}
\begin{figure}
\includegraphics[width=\linewidth,bb=0 25 800 450,clip]{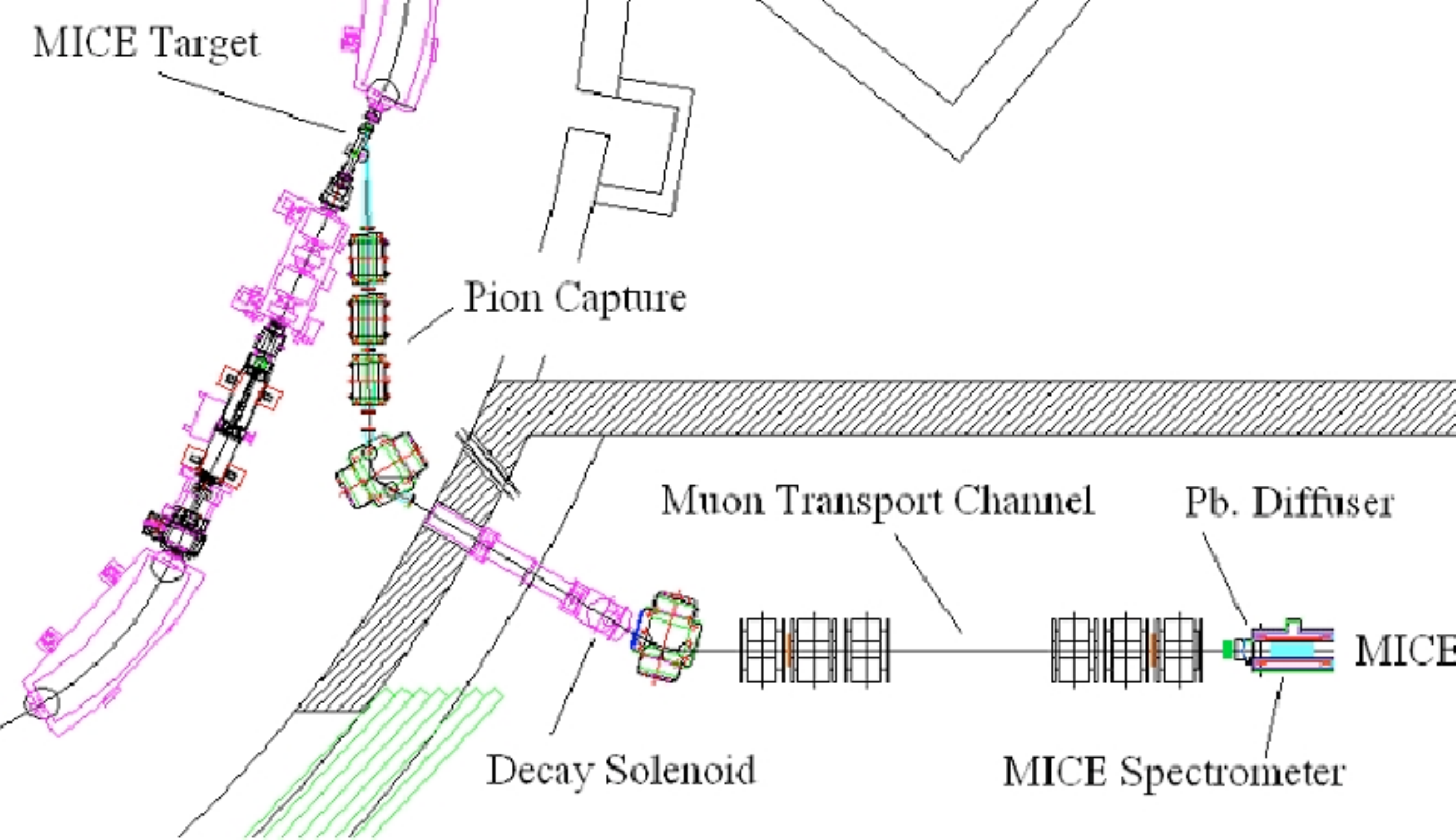}
\vspace{-.15in}
\caption{The MICE beamline at ISIS.}\label{fig:beam}
\vspace{-.08in}
\end{figure}

\noindent{\bf\it Beamline.}
The MICE beamline is sketched in Fig.~\ref{fig:beam}. Pions are produced in a small target dipped into the proton halo during the 2\,ms 800\,MeV ISIS flat-top. The target has been built (Sheffield)  and is under test. An issue with bearing wear has been found and is currently under study.

The pions are captured by a quadrupole triplet and momentum-selected by a dipole, all situated within the ISIS vault. The installation of this part of the beamline is complete. Pions decay within a 5-m-long, 12-cm-bore, 5\,T solenoid provided by PSI in Switzerland. Refurbishing and warm testing of the solenoid are complete and the cryogenic plant has been purchased. Solenoid installation is planned for later this year.

The pion-decay muons are momentum-selected in a second dipole, at a momentum about half that of the first (backward pion decay), ensuring excellent muon purity. In order to have a large momentum bite, the second bend angle is half the first one. Matching of the beam and verification of its composition are performed in the last beam section, consisting of two quadrupole triplets. 

\begin{figure}[b]
\vspace{-.08in}
\centerline{\includegraphics[width=.45\linewidth]{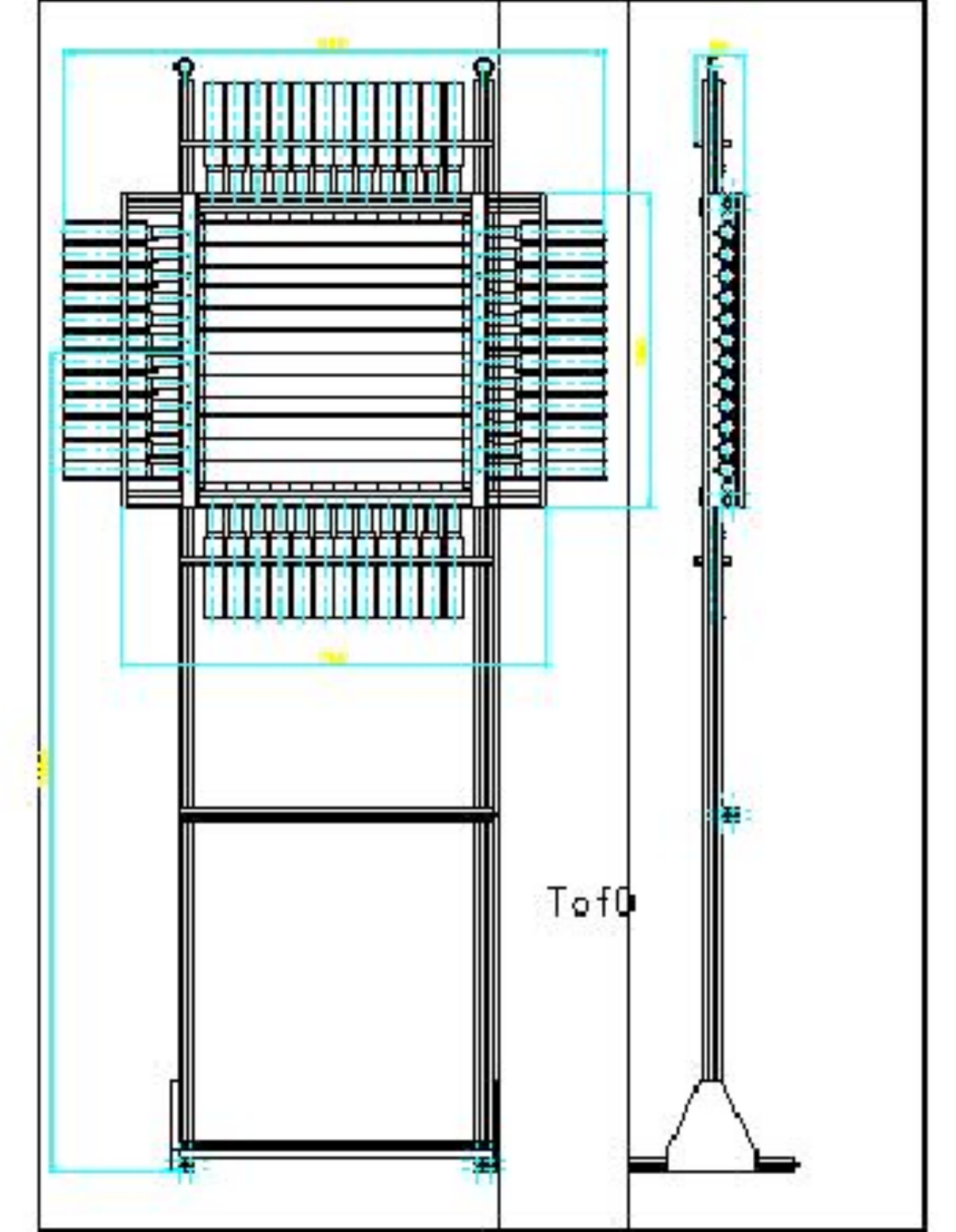}\includegraphics[width=.45\linewidth]{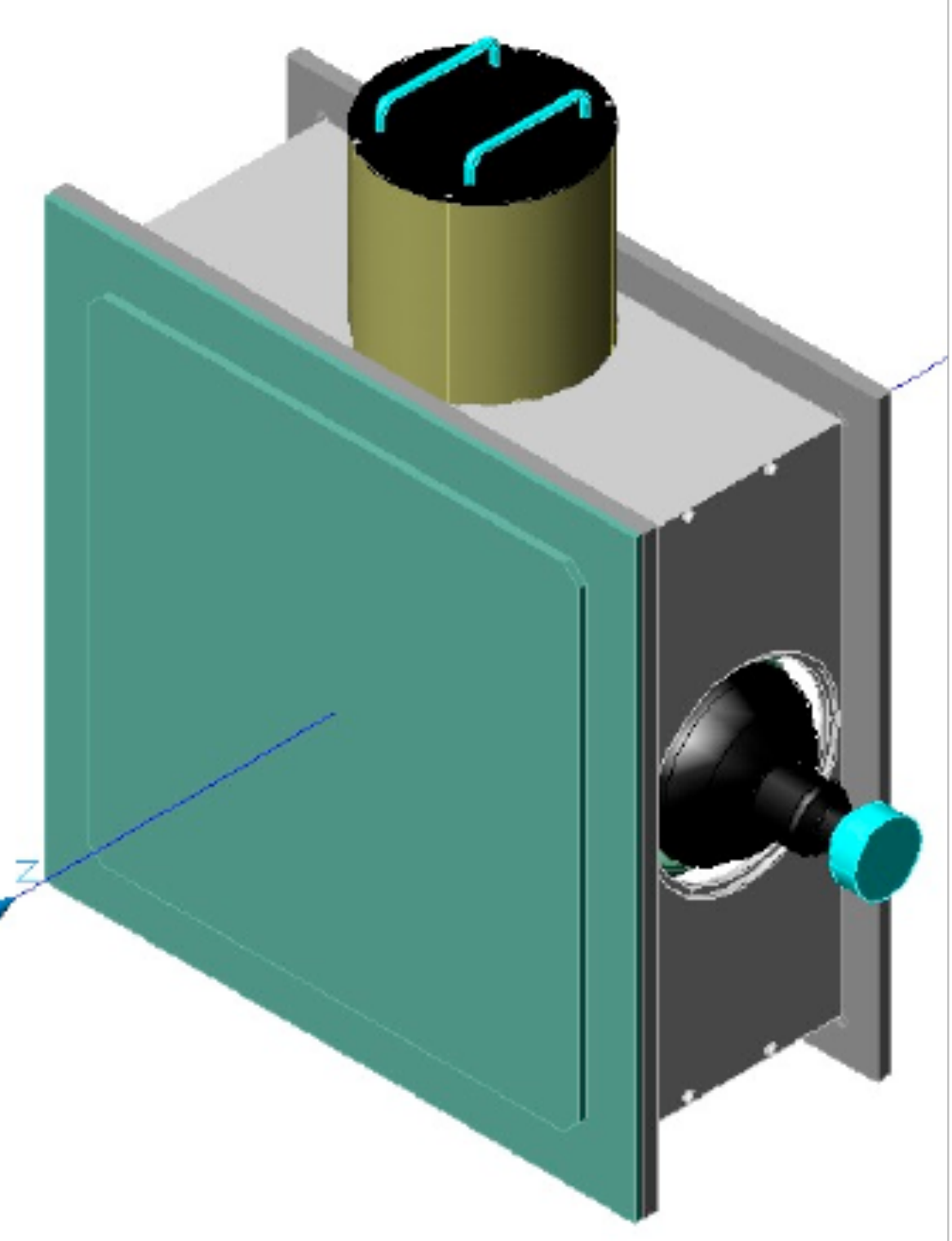}}
\caption{The upstream particle-ID detectors.}\label{fig:PID}
\end{figure}

The first TOF station, a double-layer scintillator hodoscope (Italy, with assistance from Geneva and Sofia), and the beam Cherenkov counters (Belgium-US) are situated between the two triplets (see Fig.~\ref{fig:PID}). A second TOF at the entrance of the first MICE spectrometer, with 50\,ps resolution, complemented by the Cherenkov counters, provides $\pi/\mu$ separation up to 300\,MeV/$c$. Prototype TOF and Cherenkov counters were beam-tested in summer 2006. Assembly of the TOF and Cherenkovs is in progress, with anticipated installation towards the end of this year. 

\begin{figure}[tb]
\vspace{-.05in}
\includegraphics[width=\linewidth]{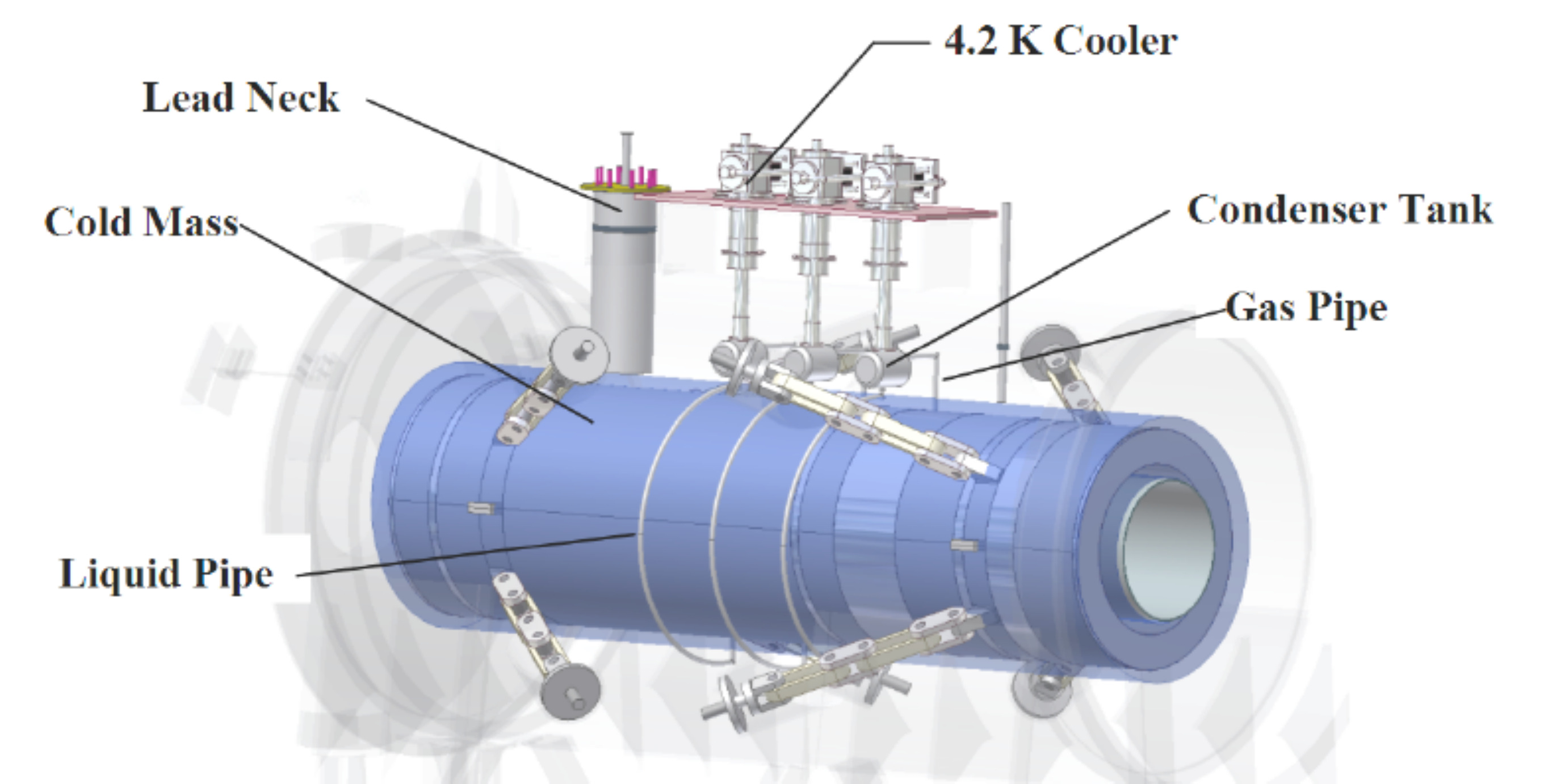}
\caption{The MICE spectrometer solenoid.}\label{fig:spect}
\end{figure}

\begin{figure}[tb]
\centerline{\includegraphics[width=.8\linewidth]{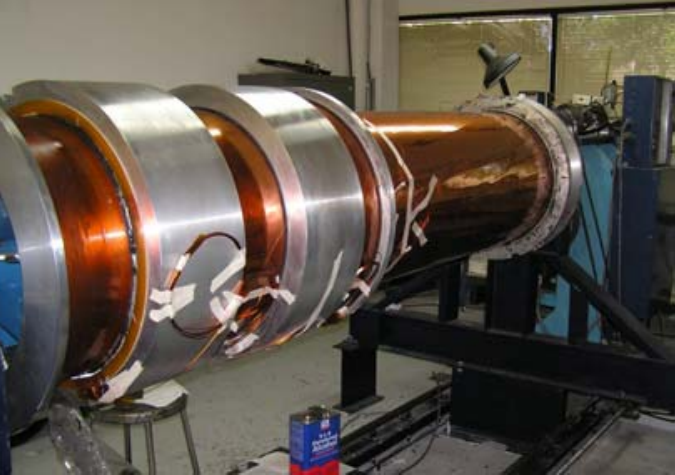}}
\caption{Spectrometer solenoid coils being wound.}\label{fig:winding}
\vspace{-.1in}
\end{figure}

\noindent{\bf\it Spectrometers.}
The 2-m-long spectrometer solenoids (US; see Figs.~\ref{fig:spect} and \ref{fig:winding}) will provide 4\,T over a 1-m-long, 20-cm-radius tracking volume. Two ``end-coils" ensure $<$1\% field non-uniformity; two additional coils at one end match optics into and out of the cooling cell. The magnets are on order for winter 2007/8 delivery, followed by measurement at Fermilab and spring 2008 installation at RAL. Magnet sensors (NIKHEF) will monitor the field during the many foreseen variations of MICE optics. 


\noindent{\bf\it Tracking detectors.}
These are in collaborative development  by Japanese, UK, and US MICE groups. Following a cosmic-ray test in 2004, a prototype was  tested with magnetic field in a beam at KEK in 2005.
The prototype showed alignment capability at the level of 5--10\,$\mu$m and 440\,$\mu$m resolution. The resolution achieved by the tracker in $x, x^\prime$ is less than 10\% of beam sizes at equilibrium emittance, ensuring unfolding of detector resolution from the emittance measurement with small systematics. Production is well advanced, with a careful QC process to avoid any loss of light yield or efficiency. The first tracker will be in place for first beam later this year. 

\begin{figure}[b]
\vspace{-.08in}
\includegraphics[width=\linewidth]{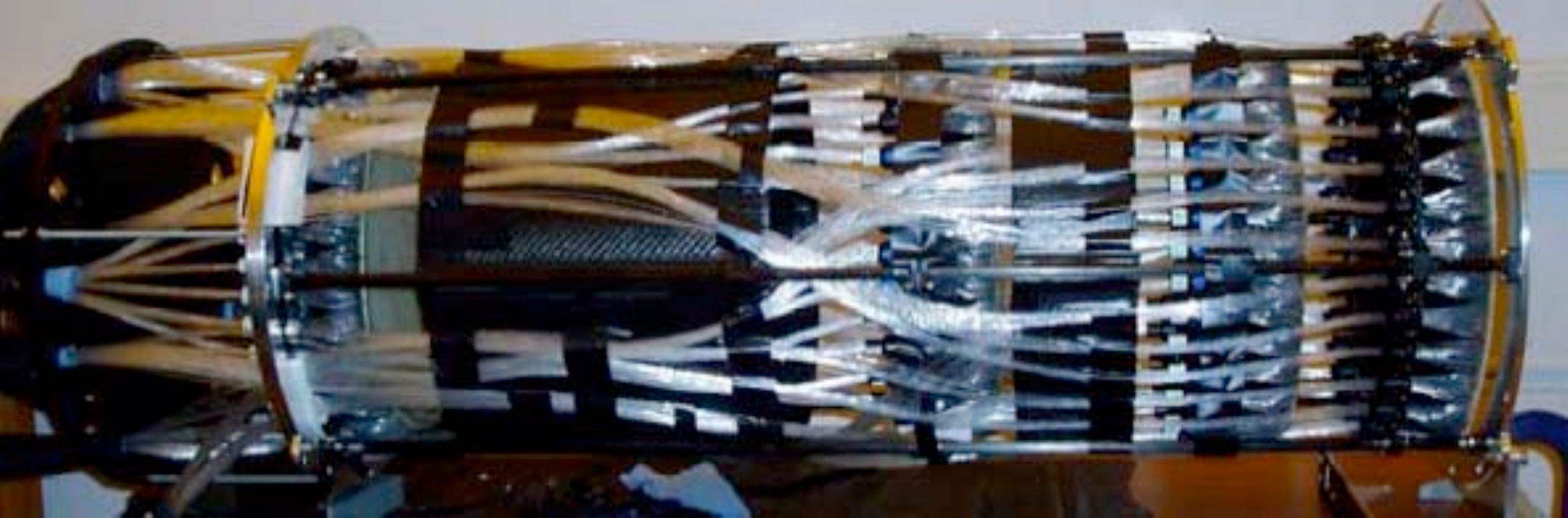}
\caption{Four-station prototype of the MICE tracker.}\label{fig:proto}
\vspace{-.08in}
\end{figure}

\noindent{\bf\it Downstream Particle ID.}
This features a TOF station and a calorimeter (Italy-Geneva-Sofia). The calorimeter separates muons from decay electrons. The design features a first layer of lead--scintillating-fiber sandwich, similar to the calorimeter of the KLOE experiment in Frascati, followed by a fully sensitive segmented scintillator block of about 1\,m$^3$. The sandwich layer degrades electrons and the scintillator makes a precise measurement of muon range. Prototypes have been tested in the Frascati BTF. Assembly of the final detectors is in progress. 

\noindent{\bf\it Controls and monitoring.}
MICE will run at $\approx$1\,Hz and record  up to 1000 muons in each 1\,ms beam burst. RF amplitude and phase, liquid-hydrogen mass, and many other parameters must be known so as to compare precisely the measured cooling effect with predictions. State-of-the-art instrumentation will be constantly monitored and relevant parameters recorded in the data stream (Daresbury-Geneva).

\noindent{\bf\it Cooling cell: absorber--focus pair.}
The absorber--focus-coil (AFC) module integrates two functions of the cooling cell. An absorber reduces the muon momentum in all dimensions, and a pair of focusing coils reduces the beta function to ensure a small equilibrium emittance. The best absorption medium is hydrogen, but for emittance far from equilibrium, a slightly less efficient but more practical medium may be He or a solid such as LiH or Be. MICE is designed to test liquid or solid absorbers for a range of beta-functions.
 
Critical issues for the hydrogen system are (1) metal containment windows as thin as possible, (2) safety, and (3) hydrogen storage. Absorber bodies and thin windows are under test in the MuCool program at KEK and Fermilab. A prototype storage system using metal-hydride beds is under development at RAL. Safety reviews, both internal and by the RAL safety office, have been passed successfully.  AFC-module construction will begin upon approval of the MICE-UK Phase II funding bid. R\&D on these critical items is well advanced.

\begin{figure}[t]
\vspace{-.05in}
\includegraphics[width=.5\linewidth]{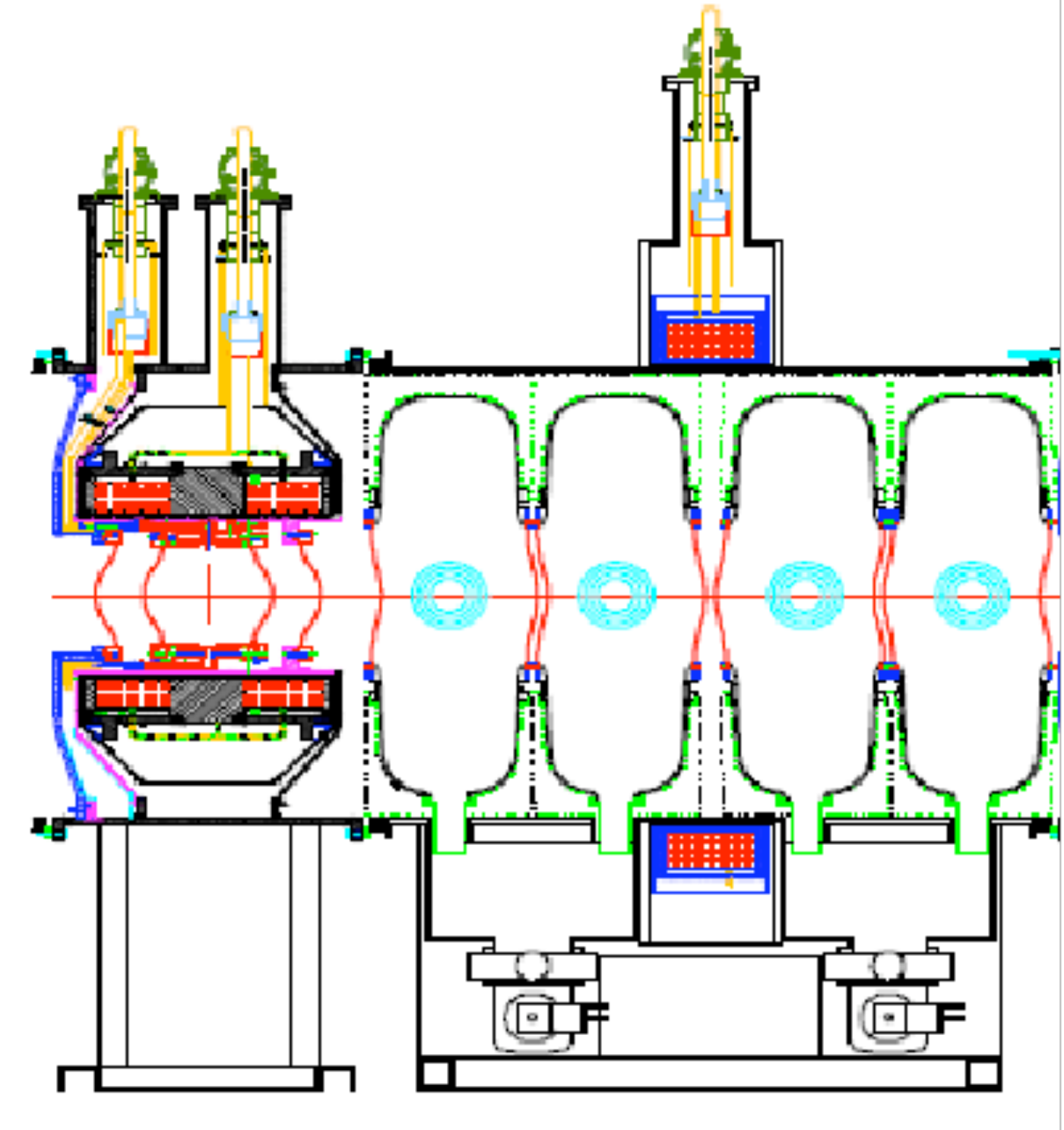}~~\includegraphics[width=.41\linewidth]{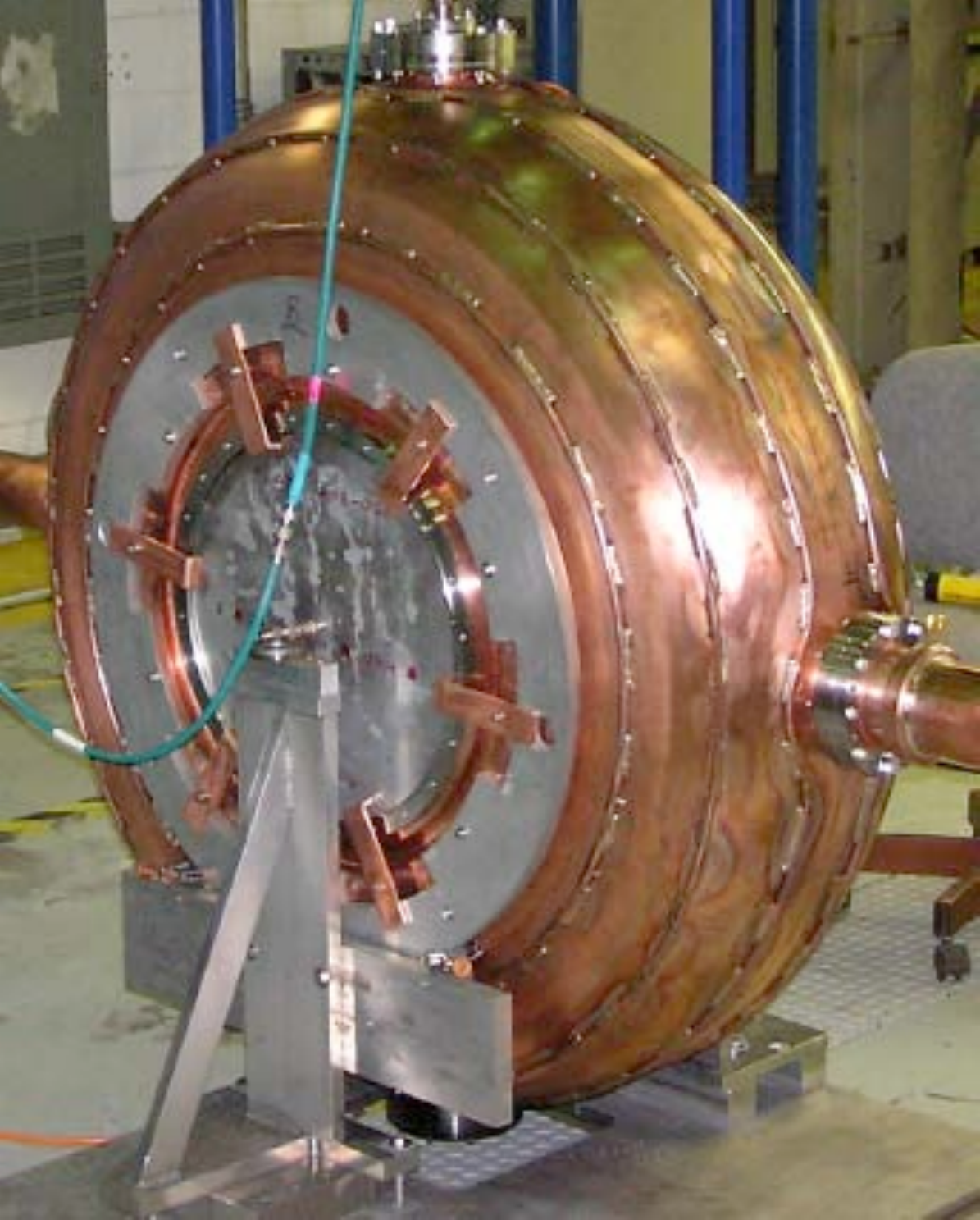}
\caption{Schematic of RFCC module connected to AFC module, with cavities closed by curved Be windows; 201-MHz cavity built by LBNL and Jlab.}\label{fig:RF}
\end{figure}

\begin{figure}[t]
\vspace{-.05in}
\centerline{\includegraphics[width=.5\linewidth]{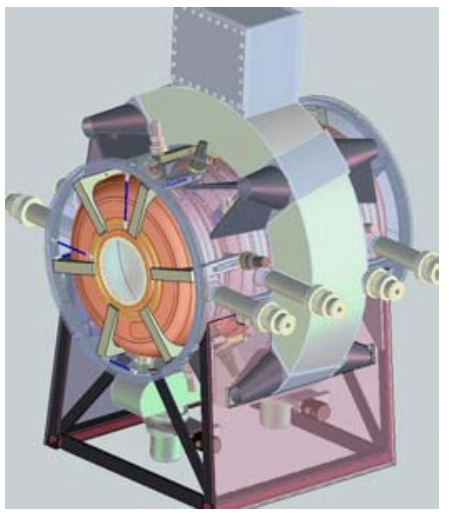}}
\caption{3D rendering of RFCC module showing cavities  with their 6-spoke tuners, pumping manifold, and input couplers.}\label{fig:RFCC}
\end{figure}

\noindent{\bf\it Cooling cell: RFCC module.}
The RF system requires relatively low frequency to handle the large beam, and must be located in a magnetic field, precluding use of SC cavities. A large ``coupling" coil surrounds the cavities. Figure~\ref{fig:RF} illustrates an RF--Coupling Coil module joined with an AFC module, together with the prototype 201\,MHz cavity  (powered at Fermilab up to 16\,MV/m without magnetic field); Fig.~\ref{fig:RFCC} shows the design in more detail. Crucial high-magnetic-field tests await provision of a Coupling Coil. RF power sources up to 4\,MW are being refurbished at  Daresbury from used systems donated by LBNL and Los Alamos, and CERN is refurbishing  another 4\,MW system, allowing MICE to run with a total acceleration of 23\,MeV in Step VI. 

 The Coupling Coil is being designed by the Institute of Cryogenics and Superconductivity Technology of the Harbin Institute of Technology (China), in collaboration with LBNL. The design is complete and the superconductor will soon be ordered.


\section{Schedule}
\begin{figure}[t]
\centerline{\includegraphics[width=\linewidth]{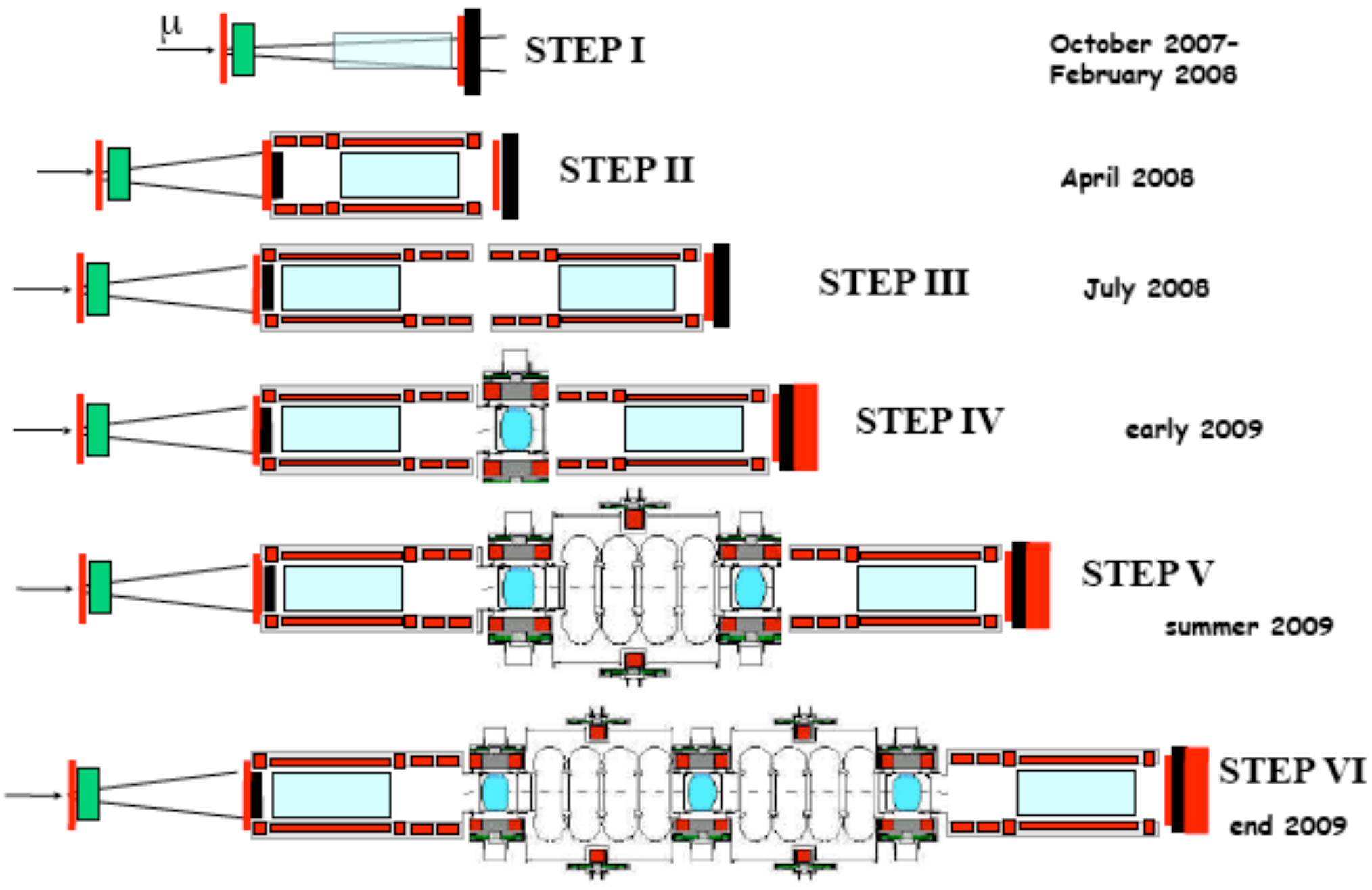}}
\vspace{-.1in}
\caption{Planned stages of MICE.}\label{fig:stages}
\end{figure}

The staged construction of MICE suits both experiment methodology and funding realities.
In the current schedule, steady beam will be available in January 2008, with occasional parasitic  availability in the preceding months. At that point the main PID detectors will be ready as well as the tracker (without magnetic field). This will allow run-in of detectors and data acquisition and characterization of the beam  (with an estimate of muon energy from the calorimeter). This ``Step I" (Fig.~\ref{fig:stages}) will also be valuable for tracker alignment.

As soon as the first spectrometer solenoid becomes available, a measurement of particle momenta, and thus of emittance, will be possible (Step II). 
Step III is crucial. By comparing directly two emittance measurements with high precision, it will allow a precise determination of the measurement biases and test the correction procedures. 

Steps I--III of MICE are fully funded. Step IV will take place, assuming timely funding of the MICE-UK ``Phase II" bid, in early 2009. A first measurement of cooling will be performed using the first AFC module. This will also allow a test of the focusing optics, both in ``flip" (the pair of focus coils oppositely powered) and non-flip modes, for values of the lattice $\beta$ function from 5 to 42\,cm.
With Step V begin tests of ``sustainable cooling," momentum lost in absorbers being restored in RF cavities. 
Finally with Step VI, a full cooling cell will be tested.

\end{document}